\documentclass[11pt]{article}
\usepackage{amsmath,amssymb,epsfig,cite,color,verbatim,array,bm,amsfonts,enumerate,enumitem}
\usepackage{graphics,float}
\usepackage[colorlinks=true,citecolor=DarkOrchid,urlcolor=black,linkcolor=Blue]{hyperref}
\hypersetup{ linktoc=page}
\usepackage{blindtext}
\usepackage{microtype}
\usepackage[latin9]{inputenc}
\usepackage{authblk}
\everymath{\displaystyle}
\usepackage{stackrel} 
\usepackage{float}
\usepackage{booktabs}
\usepackage{epsfig}
\usepackage{caption}  
\usepackage{booktabs}
\usepackage{tabularx,array}
\usepackage{siunitx}
\usepackage{tensor}
\newcolumntype{L}{>{\collectcell\AddLabel}r<{\endcollectcell}}

\newcounter{mycounter}
\usepackage[caption = false]{subfig}

\numberwithin{equation}{section}
\allowdisplaybreaks

\usepackage[dvipsnames]{xcolor}

\title{Black string spectrum of shift-symmetric Horndeski theories}
 
\author{Luis Guajardo \footnote{luis.guajardo.r-at-gmail.com } }

\affil{ {\small {\it Instituto de Investigaci\'on Interdisciplinaria, Vicerrector\'ia Acad\'emica, Universidad de Talca, 3460000 Talca, Chile} }} 

\begin{document}

\maketitle

\begin{abstract} 
In the present work, we study four-dimensional black strings in Horndeski models with translation invariance. Imposing that the scalar field depends on the string-generator coordinate, the Klein-Gordon equation admits a linear profile as a solution. This relaxation allows finding rotating, asymptotically AdS$_3\times \mathbb{R}$ black strings, dressed with an effective cosmological constant. In this regard, we show that in the full spectrum of shift-symmetric Horndeski theories with Einstein limit, the scalar charge needs to be fixed in terms of the parameter space.  This method is employed to concrete examples to illustrate the scheme we go along with. Regarding the conserved charges, we compute them via the Euclidean method and show the fulfillment of the associated Smarr Law. Finally, we exhibit that our AdS strings are locally and globally stable under small fluctuations around the equilibrium.
\end{abstract}

\section{Introduction \label{sec:Intro}}

Black strings, black rings, or $p-$branes in general, are an interesting class of solutions that enrich the phenomenology outreached by General Relativity (GR). At the fundamental level, the Emparan-Reall black ring \cite{Emparan:2001wn} illustrates that the classical uniqueness theorems \cite{Israel:1967wq, Carter:1971zc, Robinson:1975bv} are no longer valid in five or higher dimensions. Being topologically distinct from black holes, these solutions feature an \emph{extended} event horizon, in which the extra dimensions play a leading role in the existence of a linear unstable mode: the so-called Gregory-Laflamme (GL) instability \cite{Gregory:1993vy}.

Although the oxidation of the Schwarzschild black hole with extended Euclidean coordinates is trivial, at the time of introducing a cosmological constant the process gets more involved. A simple way to obtain exact black strings and $p-$branes with a negative cosmological constant in arbitrary $D=3+p$ dimensions is by introducing $p\geq 1$ massless scalar fields homogeneously distributed along the brane generator coordinates \cite{Cisterna:2017qrb}. The main idea relies on the fact that these scalar fields can be minimally coupled to gravity, and the integration constant coming from the Klein-Gordon equation (hereon we will refer to it as the \emph{scalar charge}) is fixed in terms of the cosmological constant. Surprisingly, it was shown that these strings do not trigger the GL instability \cite{Cisterna:2019scr}. 


Notwithstanding black strings were introduced as higher-dimensional objects, in four dimensions they have also provided interesting phenomena. In addition to the case $p=1$ from Ref.\cite{Cisterna:2017qrb}, the dynamic sector of Chern-Simons Modified Gravity (CSMG) \cite{Jackiw:2003pm} admits rotating and torsional black strings \cite{Cisterna:2018jsx}. CSMG is an extension of GR that can be motivated by the anomaly cancellation in curved spacetimes, and string theory \cite{Alexander:2009tp}. Since the initial value problem is not generically well-posed, the dynamical sector of CSMG should be thought of as an effective theory \cite{Delsate:2014hba}. To be self-contained, the action considered in Ref. \cite{Cisterna:2018jsx} reads:

\begin{equation}
\label{eq:action_modcs} S[g_{\mu\nu}, \phi] = \dfrac{1}{2\kappa}\int d^4x\sqrt{-g} \left[ R - 2\Lambda + \dfrac{\alpha}{4}\phi\ ^{\star}RR\right]  - \dfrac{1}{2} \int d^4x\sqrt{-g} \nabla_{\mu} \phi \nabla^{\mu} \phi.
\end{equation}
Here, $\kappa=8\pi G$, $\tensor[^\star]{R}{^\mu^\nu^\lambda^\rho}=\dfrac{1}{2}\epsilon^{\lambda\rho\sigma\tau}\tensor[]{R}{^\mu^\nu_\sigma_\tau}$ is the dual of the Riemann tensor, and $\tensor[^\star]{RR}{}$ is the Pontryagin density, defined as: 
\begin{equation}
\tensor[^\star]{RR}{} \equiv \dfrac{1}{2}\epsilon^{\gamma\delta\tau\sigma}\tensor[]{R}{^\mu_\nu_\gamma_\delta}\tensor[]{R}{^\nu_\mu_\tau_\sigma},
\end{equation}
with $\epsilon^{\gamma\delta\tau\sigma}$ the Levi-Civita tensor. At this point, we would like to stress that unless otherwise stated, we will set throughout this manuscript a stationary and axisymmetric three-dimensional base manifold with one extended Euclidean coordinate ($p=1$), that is, a metric tensor belonging to the following family:
\begin{equation}
\label{eq:ansatz} ds^2 = -N^2(r)f(r)~dt^2 + \dfrac{dr^2}{f(r)} + r^2(J(r)dt + d\theta)^2 + dz^2.
\end{equation}

The equations of motion obtained from the variation of (\ref{eq:action_modcs}) reads:
\begin{eqnarray*}
&&G_{\mu\nu} + \Lambda g_{\mu\nu} + \alpha C_{\mu\nu} = \kappa \left(\nabla_{\mu}\phi \nabla_{\nu}\phi - \dfrac{1}{2}g_{\mu\nu}\nabla_{\lambda}\phi \nabla^{\lambda}\phi \right), \\
&&\square \phi + \dfrac{\alpha}{8\kappa}\, {^{\star}RR} = 0,
\end{eqnarray*}
where $$C^{\mu\nu} = \nabla_{\rho}\phi \epsilon^{\rho\sigma\lambda (\mu}\nabla_{\lambda}R^{\nu )}_{\ \sigma} + \nabla_{\rho}\nabla_{\sigma}\phi \tensor[^\star]{R}{^\sigma^{(\mu} ^{\nu)}^\rho}.$$
As noted in \cite{Cisterna:2018jsx}, the vanishing of the Pontryagin density on (\ref{eq:ansatz}) is a valuable resource in obtaining the string solution, since the massless Klein-Gordon equation is recovered, $\square \phi =0$. If one imposes the scalar field to be dependent on the string coordinate, $\phi=\phi(z)$, a linear profile is obtained:
\begin{equation}
\label{eq:scalar_field}\phi(z) = \omega z + c,
\end{equation}
where the shift-symmetry of the field space (i.e., invariance under the map $\phi \mapsto \phi + c$) protects us to work with $c=0$ without any loss of generality. Furthermore, they showed that solving the condition $C_{\mu\nu}=0$ for the CSMG-string was enough to ensure compatibility with the full system of equations. As a consequence, the coupling constant $\alpha$ in the model (\ref{eq:action_modcs}) is in some sense untraceable for these CSMG-strings, and the solution can be thought of as an extension of the Ba\~nados-Teitelboim-Zanelli (BTZ) black hole \cite{Banados:1992wn}. As was the case for the minimal coupling, the scalar charge is fixed in terms of the cosmological constant. Recently, the thermodynamic properties of these rotating CSMG-strings, as well as stability and string-soliton phase transitions have been discussed in detail in Ref. \cite{Corral:2021tww}.

In contrast with CSMG, Horndeski's theory is a scalar-tensor extension of GR that keeps second-order dynamics for the metric and the scalar field. This ensures the theory to be healthy in terms of ghost pathologies, hence interesting for theoretical, as well as observational studies~\cite{Horndeski:1974wa, Kobayashi:2019hrl}. A concrete example is the shift-symmetric sector, defined as the subclass where the action is invariant under the mapping $\phi \mapsto \phi + c$, which was restricted by a no-hair theorem for static and spherically symmetric spacetimes \cite{Hui:2012qt}, forbidding a non-trivial scalar field. However, it was realized that hairy solutions can be constructed in spherical symmetry by bypassing some of the hypotheses \cite{Sotiriou:2013qea, Sotiriou:2014pfa, Babichev:2016rlq} (see also   \cite{Andrade:2013gsa, Baggioli:2021ejg, Cisterna:2018hzf, Cisterna:2019uek, Kobayashi:2014eva, Babichev:2013cya, Anabalon:2013oea, Bravo-Gaete:2014haa, Bravo-Gaete:2013dca, Bravo-Gaete:2021hlc}). From an astrophysical motivation, massive scalars are expected to exponentially decay at a rate proportional to the inverse of their mass. Shift-symmetric models provide massless or ultralight scalars, hence they can be effectively analyzed in strong gravity observations \cite{Thaalba:2022bnt}. Some of these hairy solutions were indeed probed against the M87* data, obtaining constraints for the scalar hair \cite{Vagnozzi:2022moj}. In terms of the modern Galileon formulation, the shift-symmetric sector of Horndeski theory can be written as~\cite{Deffayet:2011gz, Deffayet:2013lga}
\begin{equation}
\label{eq:action_horneski}S[g_{\mu\nu},\phi] = \int d^4x\sqrt{-g} \sum_{n=2}^5 \mathcal{L}_n\,,
\end{equation}
with
\begin{eqnarray*}
&&\mathcal{L}_2 = G_2(X)\,,\\
&&\mathcal{L}_3 = -G_3(X)\square \phi\,,\\
&&\mathcal{L}_4 = G_4(X)R + G_{4X}(X)\left[ (\square \phi)^2 - \nabla_{\mu}\nabla_{\nu}\phi \nabla^{\mu}\nabla^{\nu}\phi \right]\,,\\
&&\mathcal{L}_5 = -\dfrac{G_{5X}(X)}{6}\left[ (\square \phi)^3 - 3\nabla_{\mu}\nabla_{\nu}\phi \nabla^{\mu}\nabla^{\nu}\phi \square \phi + 2\nabla^{\nu}\nabla_{\mu}\phi \nabla^{\alpha}\nabla_{\nu}\phi \nabla^{\mu}\nabla_{\alpha}\phi \right] \\
&&\hspace{1cm}+G_5(X)G^{\mu\nu}\nabla_{\nu}\nabla_{\mu}\phi\,.
\end{eqnarray*}
Here, $R$ is the Ricci scalar, $G_{\mu\nu}$ is the Einstein tensor, $X = -\dfrac{1}{2}\nabla_{\mu}\phi \nabla^{\mu}\phi$ stands for the kinetic term, and the subscript $G_{X}$ denotes the derivative with respect to $X$.

In the case of black strings non-trivial profiles are allowed, but the previous models share that the scalar charge is fixed in terms of the cosmological constant, as a consequence of the compatibility between the string and the transverse sector, which is required to solve the equations. In this direction, note as an additional example that the previous results on CSMG-strings \cite{Cisterna:2018jsx, Corral:2021tww} remain valid if one swaps the Pontryagin density in (\ref{eq:action_modcs}) for the Gauss-Bonnet (GB) density,
\begin{equation}
\label{GB}\mathcal{G} = R^2 -4R_{ab}R^{ab} + R_{abcd}R^{abcd}.
\end{equation}
Indeed, the discussion is more evident in this case, since the GB density identically vanishes on (\ref{eq:ansatz}). This is due to the absence of non-vanishing z-components of the Riemann tensor, so the Gauss-Bonnet density on the four-dimensional ansatz is equivalent to its restriction to the three-dimensional base manifold. Again, the contribution of the non-minimal coupling $\phi \mathcal{G}$ will be lost and the scalar charge will be fixed in terms of $\Lambda$. Thus, it is valid to ask if this fixing issue can be considered as a general property for these black strings. In this work, we will cover Horndeski models with shift-symmetry, using the general static ansatz (\ref{eq:ansatz}) and to look for conditions that made possible to answer the previous question. If one imposes to the scalar field a pure string coordinate dependency, the shift symmetric sector is compatible with $\square \phi = 0$, and therefore, a linear profile for the scalar field is always a solution of the Klein-Gordon equation. Using this, we will show that as long as it admits an Einstein limit (defined in such a way that the Einstein-Hilbert model is recovered in the absence of the scalar field, i.e., $G_4(0)=1$), any model supporting a four-dimensional black string solution with a linear profile for the scalar field imposes an additional restriction into its charge, fixing it in terms of the parameter space.

In the attempt to organize the following information, in Sec. \ref{sec:MR} we will explicitly give the requirements that lead to a four-dimensional asymptotically AdS$_3\times \mathbb{R}$ black string, arguing why the scalar charge fixes when the theory possesses an Einstein limit. We perform a thermodynamic analysis for the black string solutions in Sec. \ref{sec:Thermo}, also analyzing their stability around the equilibrium, and showing that the thermodynamic quantities fulfill a Smarr relation. We illustrate our results in Sec. \ref{sec:example}, and finally, Sec. \ref{sec:conclusions} is devoted to our conclusions and further remarks.

\section{Main Result \label{sec:MR}}

As we stated in the introduction, the ansatz for the metric and the scalar field makes the Klein-Gordon equation compatible with $\square \phi =0$. To see this, let us start by analyzing the scalar equation for Eq. (\ref{eq:action_horneski}), given explicitly in Appendix B from Ref.~\cite{Kobayashi:2011nu}. Within our context, and for a generic function $K(X)$, some useful properties appears:
\begin{eqnarray}
\label{eq:relations_st}
&&\nabla_{\mu}X = -\nabla_{\mu}\phi \square \phi\,,\\
&&\nabla_{\mu}K(X)= -K_X\nabla_{\mu}\phi \square \phi\,,\\
&&\nabla^{\mu}\left[ K(X)\nabla_{\mu}\phi \right] = \nabla^{\mu}K(X)\nabla_{\mu}\phi + K(X)\square\phi = \square \phi (-K_X\nabla^{\mu}\phi \nabla_{\mu}\phi + K(X))\,,\\
\label{eq:relations_end}
&&\nabla_{\mu}\nabla_{\nu}\phi = \square \phi \delta_{\mu}^{z}\delta_{\nu}^z\,.
\end{eqnarray}
Since all the z-components of the Riemann tensor are zero, the mixed terms of the form $R_{\mu\nu}\nabla^{\mu}\phi$, $R_{\rho\mu\sigma\nu}\nabla^{\nu}\phi\nabla^{\alpha}\nabla^{\beta}\phi$, $\dots$ vanish. Thus, for any shift-symmetric Horndeski model, the condition $\square\phi =0$ always fulfills the Klein-Gordon equation. Hence, we can always assume a linear profile for the scalar field, as in Eq.~(\ref{eq:scalar_field}). As a consequence, the variations of $\mathcal{L}_3$ and $\mathcal{L}_5$ with respect to the metric fields vanish; any solution will be independent of the functions $G_3$ and $G_5$. Thus, one does not need to impose parity invariance ($\phi \to -\phi$), since it will be recovered anyway. All these arguments invite us to study
\begin{equation}
\label{eq:action_gral}S[g_{\mu\nu},\phi] = \int d^4x\sqrt{-g}\left( \mathcal{L}_2 + \mathcal{L}_4\right)\,,
\end{equation}
with further details.

The variation of the action (\ref{eq:action_gral}) with respect to the metric can be reduced, after inserting relations (\ref{eq:relations_st})--(\ref{eq:relations_end}), to
\begin{equation}
\label{eom} \mathcal{E}_{\mu\nu}:= G_{4}G_{\mu\nu} - \dfrac{1}{2}g_{\mu\nu}G_{2} - \dfrac{1}{2}G_{4X}R\nabla_{\mu}\phi \nabla_{\nu}\phi - \dfrac{1}{2}G_{2X}\nabla_{\mu}\phi \nabla_{\nu}\phi = 0\,.
\end{equation}
Splitting equations (\ref{eom}) to the transverse section of the string, and the string coordinate, reads
\begin{eqnarray}
\label{eq:ein_base}&&G_4G_{\mu\nu} - \dfrac{1}{2}G_2 g_{\mu\nu} =0\,,\\
\label{eq:ein_zz} &&(2XG_{4X} - G_4)R - (G_2 - 2XG_{2X}) = 0\,,
\end{eqnarray}
respectively. Note that the transverse equations are reminiscent of an Einstein-Hilbert configuration, dressed with an \emph{effective} cosmological constant $\Lambda_e$, defined as
\begin{equation}
\label{eq:lambda_e} \Lambda_e \equiv -\dfrac{G_2}{2G_4}\,.
\end{equation}
In order to obtain a black string solution, we require to match Eq. (\ref{eq:ein_base}) with (\ref{eq:ein_zz}), yielding to the following restriction:
\begin{equation}
\label{eq:rel_omega_Lambda}-\dfrac{3G_2}{G_4} = \dfrac{2XG_{2X}-G_2}{G_4 - 2XG_{4X}}\,.
\end{equation}
So far, we have: Given nonzero functions $G_2(X)$, $G_4(X)$ such that constraint (\ref{eq:rel_omega_Lambda}) is met, a four-dimensional rotating black string can be obtained. This is
\begin{equation}
ds^2 = -N^2(r)f(r)~dt^2 + \dfrac{dr^2}{f(r)} + r^2(J(r)dt + d\theta)^2 + dz^2\,,
\end{equation}
with
\begin{eqnarray}
\label{eq:metric_gral} &&N(r)=1\,,\quad J(r)= j_1 - \dfrac{j_2}{r^2}\,,\quad \phi(z)=\omega z\,, \quad f(r)= -\Lambda_e r^2 - M + \dfrac{j_2^2}{r^2}\,,
\end{eqnarray}
and $\Lambda_e$ as in (\ref{eq:lambda_e}). In this context, Eq. (\ref{eq:rel_omega_Lambda}) is fixing the scalar charge in terms of the parameter space, which may include the \emph{bare} cosmological constant $\Lambda$ and eventually some other coupling constants from particular theories. The string will be asymptotically AdS$_3\times \mathbb{R}$ if and only if in the parameter space $\dfrac{G_2}{G_4} >0$ \footnote{In the other case, when the effective cosmological constant is positive, the solution admits an extended cosmological horizon, see e.g.~\cite{Gibbons:1977mu}. In this work, we are focusing the interest on AdS-strings, and we do not further elaborate on this case.}, and it can be thought of as an extension of the BTZ black hole, dressed with an effective cosmological constant $\Lambda_e$.

There are some subtleties that we are now in position to discuss. For example, one can impose $G_4(0) = 1$ to reach an Einstein limit in the absence of the scalar field. This requirement, while natural, is double-edged: Although it avoids the ill-definition of the denominators from Eqs. (\ref{eq:lambda_e})--(\ref{eq:rel_omega_Lambda}), it induces the fixing of the scalar charge. To clarify this point let us focus on Eq. (\ref{eq:rel_omega_Lambda}), and argue by contradiction. Suppose that the scalar charge is not fixed in the parameter space, then (\ref{eq:rel_omega_Lambda}) is a differential equation that relates $G_2$ with $G_4$, in the form 
\begin{align*}
    G_2(X) \propto \dfrac{G_4(X)^3}{X}\,.
\end{align*}
Nevertheless, for $G_4(0)=1$ one obtains that $G_2(0)$ is ill-defined. This argument proves that for the scalar field profile (\ref{eq:scalar_field}), its charge is \emph{always} rigid in the full spectrum of shift-symmetric Horndeski theories with Einstein limit.

Now that we know that the scalar charge is always fixed in the parameter space, some special cases might occur. For instance, it might exists a nonzero value $X=x^{\star}$ in the parameter space satisfying $G_4(x^{\star})=0$. The equations on the base manifold (\ref{eq:ein_base}) force $G_2(x^{\star})$ to vanish, so that the transverse section is trivial. In that scenario, the equations (\ref{eq:lambda_e}) and (\ref{eq:rel_omega_Lambda}) are no longer valid, and the solution for the string can be found by solving 
\begin{equation}
\label{eq:R_xstar}R= \left. -\dfrac{G_{2X}}{G_{4X}}\right|_{X=x^{\star}}\,,
\end{equation}
as long as the RHS is well-defined. Note that in this case, one could obtain mathematically consistent, asymptotically Anti-de Sitter strings even in the absence of the kinetic term and the \emph{bare} cosmological constant~$\Lambda$. Furthermore, if $G_{2X}(x^{\star}) = 0$ and $G_{4X}(x^\star) \neq 0$, asymptotically flat strings are feasible.

On the other hand, suppose nonzero $X=x^{\star}$ such that $G_{4}(x^{\star}) \neq 0$, but $2x^{\star}G_{4X}(x^{\star})-G_{4}(x^{\star})=0$, such that the metric contribution in (\ref{eq:ein_zz}) is lost. Then, the string sector is just setting an additional constraint, which has to be satisfied in order to solve the system. In this case, one is left with (\ref{eq:ein_base}), and the following restrictions in the parameter space,
\begin{equation}
\label{eq:restriction_g2_g4}G_{4}(x^{\star}) = 2x^{\star}G_{4X}(x^{\star})\,,\quad G_{2}(x^{\star}) = 2x^{\star}G_{2X}(x^{\star})\,.
\end{equation}

\section{Black string Thermodynamics \label{sec:Thermo}}
Having the black string solutions from the previous section, we now turn to thermodynamic analysis. We will focus on our AdS strings from model (\ref{eq:action_gral}), computing the thermodynamic quantities in the Euclidean approach~\cite{Regge:1974zd,Gibbons:1976ue}. The partition function of a thermodynamic ensemble is identified with the Euclidean path integral, in the saddle point approximation around the continuation of the classical solution~\cite{Gibbons:1976ue}. In doing so, the time coordinate is periodic and its period $\beta$ is the inverse of the Hawking temperature, $\beta= T^{-1}$. The Euclidean action is then related to the Gibbs free energy $\mathfrak{G}$ by:
\begin{equation}
\label{eq:euc_Gibss} S_E = \beta \mathfrak{G} = \beta \mathcal{M} - \mathcal{S} - \beta\Omega \mathcal{J}\,, 
\end{equation}
where $\mathcal{M}$, $\mathcal{S}$ and $\mathcal{J}$ represent the mass, the entropy, and the angular momentum of the black string, respectively. In order to obtain the Gibbs free energy, in our case will be enough to consider the Wick rotating ansatz ($t\to i \tau$) of metric (\ref{eq:ansatz})~\cite{BravoGaete:2017dso}, given by
\begin{equation}
\label{eq:ansatz_euc}ds^2 = N(r)^2f(r)~d\tau^2 + \dfrac{dr^2}{f(r)} + r^2(i J(r)d\tau + d\theta)^2 + dz^2\,,
\end{equation}
together with an axionic profile for the scalar field, namely $\phi=\phi(z)$. The range for the integral will be $\tau\in~[0,\beta]$, $r\geq r_{h}$, $\theta\in [0,\sigma]$, and $z\in [0,L]$, where $r_{h}$ represents the location of the extended event horizon for the string. As was mentioned in \cite{Corral:2021tww}, shift-symmetry protects the theory for any stability concern related to the range of $z$. Working on AdS$_3\times \mathbb{R}$, the introduction of the cut-off $L$ ensures the volume of the co-dimension 2 manifold is finite, $$\iint d\theta dz = \sigma L < \infty.$$ In other words, the quantities obtained here can be effectively understood as densities. One also could consider black rings instead of black strings using the one-point compactification of the real line to study AdS$_3\times \mathbb{S}^1$, in which case $L$ is directly finite since it stands for the periodicity of the angle.

The main idea is to evaluate the action (\ref{eq:action_gral}) on the ansatz (\ref{eq:ansatz_euc}) and rewrite it in a reduced Hamiltonian form,
\begin{equation}
\label{eq:reduced_act}S_E = \int d^4x \left[ N\mathcal{H} + N^{\varphi}\mathcal{H}_{\varphi} \right] + B\,,
\end{equation}
where $B$ is a boundary term that matches the Euclidean action when the Hamiltonian constraints 
\begin{align}
    \mathcal{H}=0\,,\quad  \mathcal{H}_{\varphi}=0\,,
\end{align}
are satisfied. To fix the boundary term, one first imposes the Euclidean action to reach a minimum around the solution, $\delta S_E = 0$, requiring at the same time that the variation of the boundary term, $\delta B$, cancels all the contributions coming from the variation of the bulk action~\cite{Regge:1974zd}. In our case, the reduced Hamiltonian is given by
\begin{equation}
\label{eq:reduced_ham}\mathcal{H} = G_4f' - rG_2 - \dfrac{p^2}{2G_4r^3}\,,
\end{equation}
where $f' = \dfrac{df}{dr}$ and $p$ is the conjugate momenta of $N^{\varphi}=J(r)$, defined as
\begin{eqnarray}
\mathcal{H}_{\varphi} = p'(r) = \left(\dfrac{G_4(X)J'(r)r^3}{N(r)}\right)'\,.
\end{eqnarray}
The field equations obtained from the variation with respect to the dynamic fields $N$, $f$, $J$, $p$, $\phi$ reads
\begin{eqnarray*}
\mathcal{E}_N := \mathcal{H} = 0\,,\quad 
\mathcal{E}_f := G_4N'(r) = 0\,,\quad 
\mathcal{E}_J := \mathcal{H}_{\varphi}=0\,,\quad
\mathcal{E}_p := -\dfrac{N(r)p}{G_4r^3} - J'(r) = 0\,, 
\end{eqnarray*}
and
\begin{align*}
    \mathcal{E}_{\phi} := -\ddot{\phi}\left(G_{4X}f' - rG_{2X} + \dfrac{p^2G_{4X}}{2G_4^2r^3} \right) + \dot{\phi}^2\ddot{\phi}\left(G_{4XX}f' - rG_{2XX} -\dfrac{p^2}{2r^3}\left[\dfrac{G_{4XX}G_{4}^2 - 2G_{4X}^2 G_4}{G_4^4}\right] \nonumber \right) =0\,.
\end{align*}
Here, $\dot{\phi} = \dfrac{d\phi}{dz}$. We can readily check the consistency of these equations with the black string solution given in Eq. (\ref{eq:metric_gral}).

At this point is important to recall that $\omega$ is fixed in terms of the parameter space, as we viewed in Sec. \ref{sec:MR}. Therefore, in this work, we will consider a vanishing variation, namely $\delta \omega=0$. Using this, the variation of the boundary term $\delta B$ simply reads
\begin{equation}
\delta B = \beta \sigma L \left[-G_4(X)\delta f - J(r)\delta p \right]\Big|_{r=r_{h}}^{r=\infty}.
\end{equation}
In virtue of the field equation $\mathcal{E}_{J}=0$, one has that $p$ is constant, and we normalize it to $p = 2j_2G_{4}(X)$, just to match with the results exhibited in (\ref{eq:metric_gral}).

On the other hand, the temperature can be obtained by demanding the Euclidean metric (\ref{eq:ansatz_euc}) not having conical singularities, which in fact yields to
\begin{equation}
\label{thermo_temp}T = \dfrac{N(r_{h})f'(r_{h})}{4\pi} = -\dfrac{\Lambda_e r_{h}^4 + j_2^2}{2\pi r_{h}^3}.
\end{equation}
Using the above, the variation of the metric function is given by: $$\delta f\Big|_{r=\infty} = -\delta M,\quad \delta f\Big|_{r=r_{h}} = - \dfrac{4\pi}{\beta}\delta r_{h}.$$ Working in the grand canonical ensemble, we obtain the following boundary term:
\begin{equation}
\label{eq:bdy_term} B= \beta\sigma L G_4M - 4\pi \sigma L G_{4}r_{h}- \beta\sigma L \Omega p,
\end{equation}
where we have defined the angular velocity, $\Omega$, as 
\begin{equation}
\Omega \equiv J(\infty) - J(r_{h}) = \dfrac{j_2}{r_{h}^2}.
\end{equation}
Therefore, the comparison between (\ref{eq:euc_Gibss}) and (\ref{eq:bdy_term}) allows us to identify the thermodynamic quantities for the string. In terms of the extended horizon $r_{h}$, they are given by:
\begin{equation}
\label{thermo_mass_entropy}\mathcal{M} = G_{4}\left(-\Lambda_e r_{h}^2 + \dfrac{j_2^2}{r_{h}^2} \right) \sigma L, \quad \mathcal{S} = 4\pi G_4r_{h}\sigma L,\quad \mathcal{J} = 2G_4 j_2\sigma L,
\end{equation}
and given these results, the mass and the entropy are positive if $G_4(X)>0$.

Using (\ref{thermo_mass_entropy}) and (\ref{thermo_temp}) into (\ref{eq:euc_Gibss}), one computes the Gibbs free energy $\mathfrak{G}$ for the string:
\begin{equation}
\label{eq:Gibbs}\mathfrak{G}(T,\Omega) = \dfrac{4\pi^2G_4T^2}{\Lambda_e + \Omega^2} \sigma L.
\end{equation}
Notice that the denominator is negative to ensure a positive temperature (\ref{thermo_temp}). With this, when $G_4(X)>0$ one can check that
$$ \dfrac{\partial^2 \mathfrak{G}}{\partial T^2} \leq 0,\quad  \dfrac{\partial^2 \mathfrak{G}}{\partial \Omega^2} \leq 0,\quad  \dfrac{\partial^2 \mathfrak{G}}{\partial T^2} \dfrac{\partial^2 \mathfrak{G}}{\partial \Omega^2} - \left( \dfrac{\partial^2 \mathfrak{G}}{\partial T\partial \Omega} \right)^2  \geq 0,$$ which means that $\mathfrak{G}$ is a concave function, ensuring global stability. Furthermore, it is also interesting to study local stability, under small perturbations around the equilibrium. In our case, the specific heat capacity $c_{T}$ and the isothermal compressibility reads
\begin{eqnarray}
&&c_{T} = T\left(\dfrac{\partial \mathcal{S}}{\partial T}\right) = -\dfrac{8\pi^2 G_4 T }{\Lambda_e + \Omega^2} \sigma L \\
&&\kappa_{T} = \dfrac{1}{\mathcal{J}} \left(\dfrac{\partial \mathcal{J}}{\partial \Omega}\right) = \dfrac{\Lambda_e - 3\Omega^2}{\Omega (\Lambda_e + \Omega^2)}
\end{eqnarray}
Again, local stability under thermal fluctuations requires $G_4(X)>0$.

We end this section showing that the thermodynamic quantities (\ref{thermo_temp}), (\ref{thermo_mass_entropy}) satisfy a Smarr relation \cite{Smarr:1972kt}. To see this, we closely follow the ideas from \cite{Banados:2005hm}. It is straightforward to check that the reduced action (\ref{eq:reduced_act}), (\ref{eq:reduced_ham}) is invariant under the scalings:
\begin{eqnarray*}
&&\bar{r}= \xi r,\quad \bar{f} = \xi^2 f,\quad \bar{N}= \xi^{-2}N,\\
&&\bar{\phi} = \phi, \quad \bar{J} = \xi^{-2}J,\quad \bar{p}=\xi^2 p.
\end{eqnarray*}
From this scaling symmetry, a direct application of the N\"oether theorem shows that the quantity
\begin{equation}
C(r) = [-G_4(X)(-rf' + 2f) - J(r)(-rp'+2p)] \sigma L
\end{equation}
is conserved, $C'(r)=0$. Therefore one can evaluate the above expression at the infinity and the horizon, and $C(\infty)= C(r_{h})$ must hold. This leads to
\begin{eqnarray*}
&&C(\infty)= 2G_4(X)M \sigma L= 2\mathcal{M},\\
&&C(r_{h}) = (-G_4(X)(-4\pi Tr_{h}) - 2J(r_{h})p) \sigma L = T\mathcal{S} + 2\Omega \mathcal{J}.
\end{eqnarray*}
Hence, one gets
\begin{equation}
\label{eq:smarr} \mathcal{M} = \dfrac{1}{2}T\mathcal{S} + \Omega \mathcal{J},
\end{equation}
which is the Smarr relation.

\section{Exploring AdS black strings in Scalar-Gauss-Bonnet theories \label{sec:example}}

It is now simple to revisit, for instance, the minimally coupled case, which follows from $G_2(X)= -2\Lambda + X$ and $G_4(X)=1$. Also note that our equations (\ref{eq:ein_base}) and (\ref{eq:ein_zz}) reproduce Eqs. (10) and (11) from Ref. \cite{Cisterna:2017qrb}. 

Another simple example is the non-minimal coupling between the scalar field and the Gauss-Bonnet density, $\phi\mathcal{G}$, as we briefly mentioned in the introduction. In the Galileons formulation, this coupling is given by $G_5(X)= -4 \ln X$, but since $G_5(X)$ plays no role in the solution, we have left again with the same results from \cite{Cisterna:2017qrb}, in agreement with our hand-waving explanation.

As we have shown in Sec \ref{sec:MR}, when the Horndeski model admits an Einstein limit the scalar charge needs to be fixed in terms of the parameter space, with some special cases that might occur depending on the specific choice of the functions $G_2(X)$ and $G_4(X)$. In this section we would like to consider an explicit example to cover those scenarios. To that end, and just for simplicity, we can pick the following combination consisting in a quadratic function $G_2(X)$ and a linear function $G_4(X)$:
\begin{align*}
G_2(X) &= -2\Lambda + \eta X + 8\alpha X^2,\quad G_4(X) = 1+4\alpha X,
\end{align*} where $\Lambda, \alpha$ and $\eta$ are parameters. Recall that $G_3(X)$ and $G_5(X)$ are not needed to generate a string solution, because for the ansatz (\ref{eq:ansatz}) these terms, and their variations, vanish. Nevertheless, they can always be introduced with a physical meaning. In particular, when $G_3(X)$ and $G_5(X)$ are chosen as
\begin{align*}
G_3(X) &= -8\alpha X,\quad G_5(X) = 4\alpha \ln |X|,
\end{align*}
one realizes that the action principle reads
\begin{equation}
\label{eq:action_eta} S[g_{\mu\nu}, \phi] = \dfrac{1}{16\pi G}\int d^4x\sqrt{-g} \left[ R - 2\Lambda +\eta X - \alpha (\phi \mathcal{G} - 4G^{\mu\nu}\nabla_{\mu}\phi \nabla_{\nu} \phi + 8X\square \phi - 8 X^2) \right].
\end{equation}
These choices allow us to interpret the action (\ref{eq:action_eta}) as an extension of the minimally coupled case with a specific combination, that has been coined in the literature as Scalar-Gauss-Bonnet theories \cite{Lu:2020iav, Hennigar:2020lsl, Fernandes:2022zrq, Charmousis:2021npl, Bravo-Gaete:2022mnr} because it generalizes the five-dimensional Einstein-Gauss-Bonnet density in a four-dimensional background. This particular non-minimal interaction between the scalar field and gravity that can also be obtained by demanding conformal invariance to the Klein-Gordon equation \cite{Fernandes:2021dsb}, and as a singular limit from $D$ dimensional conformal invariant theories \cite{Babichev:2023rhn, Oliva:2011np, Giribet:2014bva}. The analysis of the action (\ref{eq:action_eta}) includes the possibility of discuss black strings in the Scalar-Gauss-Bonnet model by setting $\eta=0$, and to make contact with the minimally coupled case when $\alpha =0$.

First, note that if $\alpha >0$, then $x^{\star} = -\dfrac{1}{4\alpha}$ is a nonzero value satisfying $G_4(x^{\star})=0$. Thus, the scalar charge is fixed by $\omega^2 = \dfrac{1}{2\alpha}$. As we saw in Sec. \ref{sec:MR} we must have 
\begin{equation*}
G_2(x^{\star})=0 \implies \Lambda = \dfrac{2-\eta}{8\alpha},
\end{equation*} and consequently, the metric function is obtained by solving (\ref{eq:R_xstar}), with:
\begin{equation*}
\dfrac{G_{2X}}{G_{4X}} = \left. \dfrac{\eta + 16\alpha X}{4\alpha}\right|_{X=x^{\star}} = \dfrac{\eta - 4}{4\alpha}.
\end{equation*}
Note that in this case the equation (\ref{eq:R_xstar}) relates three independent functions, hence one expects a degenerate behavior. For example, fixing $N(r)=1$ one can cast the metric as:
\begin{eqnarray}
&&ds^2 = -f(r)~dt^2 + \dfrac{dr^2}{f(r)} + r^2(J(r)dt + d\theta)^2 + dz^2,\\
&&f(r) = \dfrac{(\eta-4)r^2}{24\alpha} + \dfrac{C_1}{r} + C_2 + \dfrac{1}{2r^2}\int r^4(J'(r))^2~dr,\quad \phi(z)=\dfrac{z}{\sqrt{2\alpha}},
\end{eqnarray}
Here $C_1, C_2$ are integration constants coming from the equation (\ref{eq:R_xstar}), and $J(r)$ is a free function. One can note that for $\eta=4$, the static case admits asymptotically flat solutions provided $\alpha >0$, followed by a time scaling. We highlight that these solutions are degenerate because the \emph{on-shell} action vanishes at the point $X=x^{\star}$, therefore they cannot be seen either as an extension of the BTZ black hole or Schwarzschild black holes. For this reason, we do not further elaborate on this particular solution.

Aside of the mentioned case, applying the constraint (\ref{eq:rel_omega_Lambda}) for the model (\ref{eq:action_eta}) fixes the scalar charge $\omega$ in terms of $\Lambda, \alpha, \eta$ via a fourth order algebraic equation:

\begin{equation}
\label{eq:omega_general} \alpha(6-\eta)\omega^4 - (8\alpha\Lambda + \eta)\omega^2 - 2\Lambda = 0,
\end{equation}
which can be solved explicitly:
\begin{equation}
\label{eq:omega_sol}\omega = \pm \left(\dfrac{(8\alpha\Lambda + \eta) \pm \sqrt{(8\alpha\Lambda + \eta)^2 + 8\alpha\Lambda(6-\eta)}}{2\alpha(6-\eta)}\right)^{1/2}.
\end{equation}

Therefore, and using $\Lambda_e$ as defined in (\ref{eq:lambda_e}), the metric for the black string in this theory can be cast as (\ref{eq:metric_gral}):
\begin{eqnarray*}
&&ds^2 = -N^2(r)f(r)~dt^2 + \dfrac{dr^2}{f(r)} + r^2(J(r)dt + d\theta)^2 + dz^2,\\
&&N(r)=1,\quad f(r)= -\Lambda_e r^2 - M + \dfrac{j_2^2}{r^2},
\quad J(r)= j_1 - \dfrac{j_2}{r^2},\quad \phi(z)=\omega z,
\end{eqnarray*}
and 
\begin{equation}
\label{eq:lambda_e_eta}
\Lambda_e = -\dfrac{\omega^2(\eta -8\alpha\omega^2)}{4(1+4\alpha\omega^2)},
\end{equation}
where $\omega$ is defined in (\ref{eq:omega_sol}). The expression for $\Lambda_e$ in terms of $\alpha, \eta$ and $\Lambda$ is not illuminating, so we have decided not to display it. The black string is asymptotically AdS$_3\times \mathbb{R}$ as long as $\Lambda_e<0$, with an effective AdS$_3$ curvature radius given by
\begin{equation}
\label{eq:ads_curvature} \dfrac{1}{\ell} \equiv -\Lambda_e.
\end{equation}

Before starting the analysis of the scalar charge in the parameter space, and subsequently the black string solution, let us briefly comment some general aspects. Note in the first place that the parameter space contains three continuous parameters $(\alpha, \Lambda, \eta)$, and we have imposed by hand $\eta \geq 0$ to avoid phantom scalars. In the following, we will further restrict ourselves to $\eta \in \{0,1,6\}$ because we can keep track of the phenomenology in the absence of the minimal coupling ($\eta=0$), the standard case $\eta=1$, and the ill-definition in the denominator in (\ref{eq:omega_sol}). Of course, other cases might be studied as well, but they don't provide further insights. For the same reasons, the case $\eta=6$ is left as an Appendix \ref{sec:eta_6}.

Secondly, the smooth limit $\alpha \to 0$ in (\ref{eq:omega_general}) recovers the minimally coupled case from Ref. \cite{Cisterna:2017qrb}. One can directly check in virtue of (\ref{eq:omega_general}) that the kinetic term was mandatory to avoid the vanishing of the cosmological constant $\Lambda$. We take this fact as an input to split the rest of the analysis depending on the sign of $\alpha$.

Thirdly, although the action (\ref{eq:action_eta}) does not possess parity invariance ($\phi \to -\phi$), as we mentioned in Sec. \ref{sec:MR} it's recovered \emph{on-shell}, thus the analysis can be carried out independent of the outer sign of $\omega$. We will present some plots using the positive outer sign just for convenience. However, some differences are likely to appear because of the internal $\pm$ sign in (\ref{eq:omega_sol}). We will refer to them as the \emph{positive} and \emph{negative} branches of $\omega$, hoping that these nicknames can be illustrative of the different outcomes due to this sign.

Last but not least, since the case $G_{4}(x^{\star})=0$ was discussed earlier, until the end of this section we will assume $G_4(X) = 1+4\alpha X \neq 0,\forall X$.

\subsection{$\alpha >0$ \label{sec:alpha_pos}}

First, let's take a look in the $\eta=0$ case, which means that the minimal coupling is absent. Regarding to the domain constraints imposed by (\ref{eq:omega_sol}), $\Lambda$ has to be restricted to non-negative values, and the negative branch of $\omega$ is forbidden due to mathematical inconsistencies. Moreover, it follows from (\ref{eq:lambda_e_eta}) that $\Lambda_e$ is positive, thus cosmological horizons appear.

The standard coupling $\eta=1$ is more interesting. First, both branches are allowed when $\Lambda$ belongs to the interval 
\begin{equation}
\label{eq:interval}I=\left[ \dfrac{-7+3\sqrt{5}}{16\alpha}\ ,\ 0\right].
\end{equation}
The negative branch ceases to exists outside this interval, in contrast with its positive counterpart (see Fig. \ref{fig:apos_omega} below). When $\Lambda$ attains its minimum on $I$, both branches fix the scalar charge to the same value, $\omega^2 = \dfrac{3\sqrt{5}-5}{20\alpha}$. From this results, it is observed that the coupling constant $\alpha$ is in some sense a scale factor, so changing its value will not modify the conclusions. In the next figure, we illustrate both branches of $\omega$ in terms of $\Lambda$:

\begin{figure}[H]
  \centering
    \includegraphics[scale=0.7]{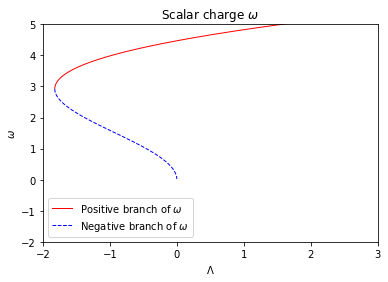}    
      \caption{\label{fig:apos_omega} $\omega$ in terms of $\Lambda$, for $\alpha>0$. The red plot (solid) corresponds to the positive branch of $\omega$, that is, considering the $+$ sign in front of the inner square root in (\ref{eq:omega_sol}), and in blue (dashed) we plot its negative branch. Note that, for a given $\alpha>0$ and $\Lambda \in I$, two possible values of $\omega$ are allowed. Since $\alpha$ is a scale factor, here we have set $\alpha=0.01$ for plotting convenience.}
\end{figure}

In virtue of the Eq. (\ref{eq:lambda_e_eta}), the effective cosmological constant has a zero at
\begin{eqnarray}
\omega^2 = 0,\quad \wedge \quad \omega^2 = \dfrac{1}{8\alpha},
\end{eqnarray}
but $\omega=0$ turns off the scalar field. Sticking to non vanishing values of $\omega$, AdS-strings are obtained for the minimally coupled case when 
\begin{equation}
\label{eq:apos_rest_omega}\omega^2 < \dfrac{1}{8\alpha},
\end{equation}
range that agrees with $G_4(X)>0$ because $$1+4\alpha X>0 \Longleftrightarrow \omega^2 < \dfrac{1}{2\alpha},$$ therefore, from the thermodynamic analysis performed in Sec. \ref{sec:Thermo} we can ensure thermodynamic stability for these AdS-strings. As we see in Fig.\ref{fig:lambdae_pos} below, the negative branch always provides AdS-strings for all $\Lambda \in I$, while the positive branch do the same but in a restricted domain: 
\begin{equation}
\label{eq:apos_pt} \dfrac{-7+3\sqrt{5}}{16\alpha} \leq \Lambda <  -\dfrac{1}{64\alpha}
\end{equation}

\begin{figure}[H]
  \centering
    \includegraphics[scale=0.7]{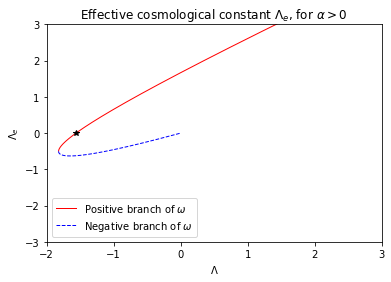}   
      \caption{\label{fig:lambdae_pos} $\Lambda_e$ in terms of $\Lambda$ for positive values of $\alpha$ and $\eta=1$. The black star is located at $\left(-\frac{1}{64\alpha},0\right)$, where the effective cosmological constant is zero. Recall that for $\eta=0$, no event horizons appear.}
\end{figure}

From the previous discussion, and as it can be seen in the figure above, when $\Lambda$ satisfies (\ref{eq:apos_pt}) both branches dress the string with an AdS asymptotic. A natural question is which solution is more stable. To answer this, one can compute their Gibbs free energies, putting the strings at the same temperature $T$ and angular velocity $\Omega$ \cite{Hawking:1982dh}. The branch with lower Gibbs energy will be preferred. For simplicity, we analyze the static case, $\Omega=0$. In the general case, one should consider the restriction $\Lambda_e + \Omega^2<0$ in (\ref{eq:Gibbs}) that ensures a positive temperature (\ref{thermo_temp}). This induces additional constraints in $\Lambda$ and $\alpha$, but no new insights.

Let's work out these ideas. In the first place, we recast the generic expression for the Gibbs free energy (\ref{eq:Gibbs}) in terms of $\Lambda_e$ and the extended horizon $r_h$:
\begin{equation}
\mathfrak{G} = G_4(X)\Lambda_e r_h^2.
\end{equation}
Bathing both strings at the same temperature induces $$\Lambda_e^{+}r_h^{+} = \Lambda_{e}^{-}r_{h}^{-},$$
(the superscript refers to the branch used) which leads to 
\begin{equation}
\label{eq:dGibbs}\Delta \mathfrak{G} \equiv \mathfrak{G}^{+} - \mathfrak{G}^{-} =  \Lambda_e^{+}(r_{h}^{+})^2 \left( G_4(X^{+}) - G_4(X^{-})\dfrac{\Lambda_e^{+}}{\Lambda_e^{-}} \right),
\end{equation}

The evaluation of the difference using both $\omega-$branches (\ref{eq:omega_sol}) and their respective cosmological constants $\Lambda_{e}^{\pm}$ (\ref{eq:lambda_e_eta}) gives as a result that the positive branch is preferred, as we can see in the next figure:

\begin{figure}[H]
  \centering
    \includegraphics[scale=0.7]{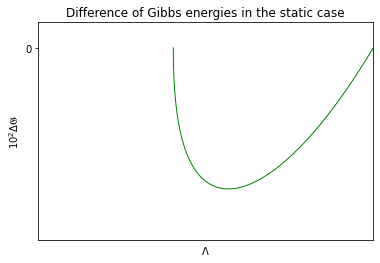}   
      \caption{\label{fig:dGibbs}Difference between the Gibbs energies computed for the positive and negative branches. In this plot, we have set $\alpha=1$ and scaled the vertical axis by a $10^2$ factor. The horizontal axis runs over the range (\ref{eq:apos_pt}), where both branches dress the string with an AdS asymptotic.}
\end{figure}

Recall that when $\Lambda$ attain its minimum, the scalar charge (and therefore $\Lambda_e$) has the same value on both branches (c.f., Fig. \ref{fig:apos_omega}). It follows that the difference between both Gibbs free energies will vanish. This corresponds to the zero on the left of the plot. On the other corner lies $\Lambda= -\dfrac{1}{64\alpha}$, in which case $\Lambda_e^{+}=0$.

\subsection{$\alpha <0$ \label{sec:alpha_neg}}

For negative values of $\alpha$, the domain for the positive branch is given by:
$$
\begin{cases}
\Lambda \geq -\dfrac{3}{4\alpha},\quad \text{when } \eta=0,\\
\ \\
\Lambda \geq -\dfrac{7+3\sqrt{5}}{16\alpha},\quad \text{when } \eta=1,\\
\end{cases}
$$
In this region, both branches coexist. Nevertheless, the negative branch is also allowed for negative values of $\Lambda$, which is mathematically inconsistent for its positive counterpart. In the next figure, a typical plot representing the values of the scalar charge in terms of $\Lambda$:

\begin{figure}[H]
  \centering
    \includegraphics[scale=0.7]{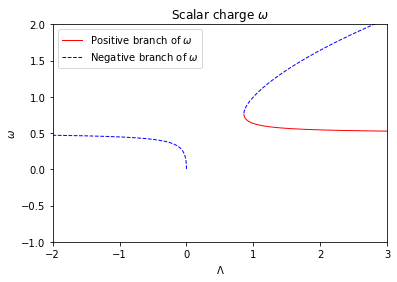}   
      \caption{\label{fig:aneg_nokin} $\omega$ in terms of $\Lambda$ for negative values of $\alpha$. The positive branch is not allowed for negative values of $\Lambda$, as expected from (\ref{eq:omega_sol}).}
\end{figure}

AdS-strings are feasible only in the negative branch. Note that thermodynamic stability is ensured since $G_4(X)=1+4\alpha X$ is always positive when $\alpha<0$. We further notice that for $\eta=0$, the value $\Lambda_e$ is equal for both cases. In turn, evaluating $\Lambda_e$ one gets: $$\Lambda_{e}^{+} - \Lambda_e^{-} = \dfrac{\eta \sqrt{(8\alpha\Lambda+\eta)^2+8\alpha\Lambda(6-\eta)}}{3\alpha(6-\eta)},$$ which clearly vanishes when $\eta=0$. The representative plot to visualize the above is the following:

\begin{figure}[H]
  \centering
    \includegraphics[scale=0.7]{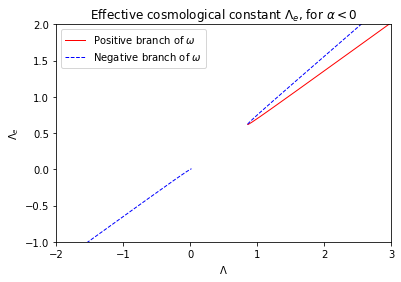}   
      \caption{\label{fig:lambdae_neg} $\Lambda_e$ in terms of $\Lambda$ for negative values of $\alpha$. Only the negative branch admits asymptotically AdS-strings, for $\Lambda <0$.}
\end{figure}

We end this section observing that in this region the value $x^{\star}=\dfrac{1}{4\alpha}$ is such that $G_4(x^\star) = 2$, but $G_4(x^{\star}) - 2x^{\star}G_{4X}(x^{\star}) =0$. As we saw in the Sec. \ref{sec:MR}, the string equation imposes a further constraint on $G_2$ (\ref{eq:restriction_g2_g4}), namely:

\begin{equation}
G_2(x^{\star}) - 2x^{\star}G_{2X}(x^{\star}) =0 \implies \Lambda = -\dfrac{\eta+6}{8\alpha}.
\end{equation}

Hence, applying these constraints to the effective cosmological constant (\ref{eq:lambda_e}), one gets

$$\Lambda_e = -\dfrac{G_2(x^{\star})}{2G_4(x^{\star})} = -\dfrac{\eta+4}{8\alpha} >0,$$
obtaining a cosmological horizon.

\section{Discussion and further comments \label{sec:conclusions} }

In summary, in this work, we have explored four-dimensional black strings in the framework of Horndeski models with shift-invariance. Imposing that the scalar field depends only on the coordinate that spans the Euclidean extension, rotating and asymptotically AdS$_3\times \mathbb{R}$ black strings in the full shift-symmetric spectrum were found, generalizing prior results with minimal couplings \cite{Cisterna:2017qrb}. In this regard, the string still is an extension of the BTZ black hole, but dressed with an effective cosmological constant. We also have shown that the existence of AdS-strings in the absence of the \emph{bare} cosmological constant is technically possible. The fixing of the scalar charge in the parameter space is possible as long as the theory admits an Einstein limit (i.e., $G_4(0)=1$ in terms of the Galileon formulation of Horndeski theory \cite{Deffayet:2011gz, Deffayet:2013lga}). If that is not the case, one could have mathematically consistent hairy scenarios. For instance, the choice $G_2(X)=-2\Lambda G_4(X)=-2\Lambda \sqrt{-X}$ satisfies the string equation trivially (\ref{eq:ein_zz}), and therefore the scalar charge is no longer fixed. 

Planar AdS black holes admit a soliton counterpart, conjectured to be the lowest energy solutions, and therefore they can be naturally chosen as the ground state of the theory \cite{Horowitz:1998ha}. In that case, phase transitions for the planar AdS black hole can occur \cite{Surya:2001vj}. In the general case, our solution (\ref{eq:metric_gral}) also admits a soliton-like counterpart, which can be obtained by a double analytic continuation ($t\mapsto i\chi, \theta \mapsto i\tau$) on the black string metric. Thus, one can also explore if phase transitions are possible in this case. Focusing on the static case for simplicity, the metric for the soliton reads
\begin{equation}
ds^{2}_{(s)} = -\rho^2 d\tau^2 + \dfrac{d\rho^2}{g(\rho)} + g(\rho)d\chi^2 + dz^2,
\end{equation}
where $g(\rho) = -\Lambda_e \rho^2 - \mu$, and $\mu$ is an integration constant. For both solutions to share the same asymptotic geometry and behavior, one imposes matching conditions at both boundary geometries, demanding periodicity in a suitable manner \cite{Anabalon:2022ksf}. For instance, notice that in contrast with the black string, the soliton coordinate $\chi$ needs to be periodic to get rid of conical singularities. Its period, $\sigma_{(s)}$, must be $$\sigma_{(s)} = \dfrac{4\pi}{g'(\rho_0)},$$ where $\rho_0$ is the largest root of $g(\rho)=0$, which induces an extra ingredient in the asymptotic geometry of the string. When the matching conditions are fulfilled, computing the difference between the Gibbs energies $\Delta \mathfrak{G} \equiv \mathfrak{G} - \mathfrak{G}_0$ directly leads to a phase transition at a critical temperature $$T_c= \dfrac{\sqrt{-\Lambda_e}}{2\pi}.$$ Our results are similar to those of \cite{Corral:2021tww}: a first-order phase transition is supported by the existence of the scalar field, the soliton is preferred when $T<T_c$ and the string in the other case, $T>T_c$. 

We have also explored a concrete example using a quadratic choice for $G_2(X)$ (c.f. (\ref{eq:action_eta})). This allows us to find a range for $\Lambda$ in which two different values for the scalar charge are possible, but no phase transitions between those configurations were observed. The difference between the Gibbs free energies vanishes on the boundary of the common interval (\ref{eq:apos_pt}), and one configuration is always preferred in its interior.

In this work, we have centered the attention on the four-dimensional scenario because Horndeski guarantees to be the most general scalar-tensor theory there. Very recently, a no-hair theorem was proven for asymptotically flat and circular black holes in any shift-symmetric model, including higher derivative extensions of Horndeski models \cite{Capuano:2023yyh}, generalizing similar results from Galileon's theory \cite{Hui:2012qt}. Unfortunately, lifting our ideas to higher dimensions is not an immediate task. Starting from the known fact that there is no analog of the Horndeski result, higher order gravitational extensions are also allowed since the Gauss-Bonnet density (\ref{GB}) is no longer a total derivative, and its dynamics affect the equations (see, e.g., \cite{Cisterna:2018mww, Cisterna:2020kde, Canfora:2021ttl, Cisterna:2021ckn}). For the same reasons, the coupling $\phi\mathcal{G}$ loses translation symmetry in five dimensions, therefore the simple idea of casting a 4D shift-invariant model in five dimensions might have non-expected consequences. Our toy model (\ref{eq:action_eta}) provides an interesting example: casting (\ref{eq:action_eta}) in $D=5$ (using as a base manifold a four-dimensional maximally symmetric spacetime and one extended coordinate), when the scalar field linearly depends on the string coordinate the matching between the transverse sector and the string coordinate is only consistent with the extension of flat space on the sphere. Nevertheless, if one restricts the problem just to analyze (\ref{eq:action_gral}) as a particular case, one can go to arbitrary $D=d+1$ dimensions, taking a $d-$dimensional Einstein manifold, and one extended coordinate. 

For the same reasons as in the previous paragraph, exploring our toy model (\ref{eq:action_eta}) in five dimensions also constitutes an interesting route to extend this work. Not only the mass parameter is lost in the attempt to match the transverse section with the string equation, but since $\phi\mathcal{G}$ is non-trivial, the latter possesses a $z$ dependency that fixes the parameter space in such a way that $\Lambda_e=0$. Employing some known techniques to endow a mass parameter into the above extension would be an interesting problem. A valid option is to try with the introduction of additional sources, such as non-linear electrodynamics \cite{Plebanski:1970zz} (see \cite{Sorokin:2021tge} for a recent review), that have been widely explored as an excellent laboratory to provide a charge-to-mass ratio that circumvents massless situations \cite{Bravo-Gaete:2022mnr, Alvarez:2022upr, Bravo-Gaete:2021hza}. These incompatibilities result from the scalar field's linearity and open a window to look for non-linear profiles.

The existence of linear unstable modes can also be a follow-up question for these strings. The minimally coupled case was proven to be stable \cite{Cisterna:2019scr}, but in the general case, the answer is not clear. Even if one sticks to recast the action (\ref{eq:action_gral}) in five dimensions, the Klein-Gordon equation now considers arbitrary functions $G_2(X), G_4(X)$ that might trigger an instability. In that sense, some additional restrictions might help to stabilize these strings in the general setup.

\section*{Acknowledgements \centering}
 
LG would like to thank the anonymous referee for the constructive remarks about a previous version of this work. Also, to M. Bravo-Gaete and K. Lara for their useful comments. Finally, LG would like to thank to Instituto de Matem\'aticas (INSTMAT), Universidad de Talca, for its hospitality during the preparation of this manuscript.

\appendix

\section{Appendix: The $\eta =6$ case \label{sec:eta_6}}

As it can be seen from (\ref{eq:omega_general}) when $\eta=6$ the scalar $X$ is fixed in terms of a linear equation instead of a quadratic one (or equivalently, the scalar charge is fixed via a quadratic equation instead of a fourth-order one). This leads to \begin{equation}
\omega^2 = \dfrac{-\Lambda}{4\alpha\Lambda + 3},
\end{equation}
whose solution exists when the RHS is non-negative. Hence, for non-vanishing values of $\Lambda$, the RHS is positive for
\begin{equation}
\label{eq:eta_6_alpha}
\alpha < -\dfrac{3}{4\Lambda}.
\end{equation}   

Computing $\Lambda_e$ from (\ref{eq:lambda_e}), one gets
\begin{equation}
\Lambda_e = \dfrac{(16\alpha\Lambda + 9)\Lambda}{6(4\alpha\Lambda+3)},
\end{equation} therefore, the effective cosmological constant is negative when
$$\begin{cases}
-\dfrac{3}{4\Lambda}<\alpha  < -\dfrac{9}{16\Lambda},\quad \text{if }\Lambda >0 \\
\ \\
\alpha  < -\dfrac{9}{16\Lambda}\,\ \vee\,\ \alpha>-\dfrac{3}{4\Lambda},\quad \text{if }\Lambda <0 \\
\end{cases}
$$

But, at the time of making it consistent with (\ref{eq:eta_6_alpha}), one observes that AdS-strings are not possible for $\Lambda>0$, and restricts $\alpha$ to $$
\alpha < -\dfrac{3}{4\Lambda},$$ when $\Lambda$ is negative.


\end{document}